\documentclass[prl,reprint,superscriptaddress,10pt]{revtex4-2}
\usepackage{amsmath}
\usepackage{hyperref}
\usepackage{graphicx}
\usepackage{xcolor}
\usepackage{epstopdf}
\usepackage{verbatim}
\usepackage{physics}
\usepackage{pgfplots}
\usepackage[normalem]{ulem}
\usepackage{siunitx}
\usepackage{orcidlink}

\begin{document}
\author{Marion Mallweger~\orcidlink{0009-0009-8142-2988}}
\email[]{marion.mallweger@fysik.su.se}
\thanks{\quad These two authors contributed equally.}
\affiliation{Department of Physics, Stockholm University, SE-106 91 Stockholm, Sweden}

\author{Milena Guevara-Bertsch~\orcidlink{0000-0003-3771-6531}}
\email[]{Milena.Guevara-Bertsch@uibk.ac.at}
\thanks{\quad These two authors contributed equally.}
\affiliation{Institut f\"{u}r Experimentalphysik, Universit\"{a}t Innsbruck, Technikerstr. 25, 6020 Innsbruck, Austria}
\affiliation{Institut f\"{u}r Quantenoptik und Quanteninformation,\"{O}sterreichische Akademie der Wissenschaften, Technikerstr. 21a, 6020-Innsbruck, Austria}

\author{Boyan T. Torosov~\orcidlink{0000-0002-6690-1712}}
\affiliation{Institute of Solid State Physics, Bulgarian Academy of Sciences, 72 Tsarigradsko chauss\'{e}e, 1784 Sofia, Bulgaria}

\author{Robin Thomm~\orcidlink{0009-0000-8105-8690}}
\affiliation{Department of Physics, Stockholm University, SE-106 91 Stockholm, Sweden}

\author{Natalia Kuk}
\affiliation{Department of Physics, Stockholm University, SE-106 91 Stockholm, Sweden}

\author{Harry Parke~\orcidlink{0000-0001-6120-5470}}
\affiliation{Department of Physics, Stockholm University, SE-106 91 Stockholm, Sweden}

\author{Christan F. Roos~\orcidlink{0000-0001-7121-8259}}
\affiliation{Institut f\"{u}r Experimentalphysik, Universit\"{a}t Innsbruck, Technikerstr. 25, 6020 Innsbruck, Austria}
\affiliation{Institut f\"{u}r Quantenoptik und Quanteninformation,\"{O}sterreichische Akademie der Wissenschaften, Technikerstr. 21a, 6020-Innsbruck, Austria}

\author{Gerard Higgins~\orcidlink{0000-0003-0946-8067}}
\affiliation{Department of Physics, Stockholm University, SE-106 91 Stockholm, Sweden}
\affiliation{Department of Microtechnology and Nanoscience (MC2), Chalmers University of Technology, SE-412 96 Gothenburg, Sweden}
\affiliation{Institut f\"{u}r Quantenoptik und Quanteninformation (IQOQI), \"{O}sterreichische Akademie der Wissenschaften, Boltzmanngasse 3, A-1090 Vienna, Austria}

\author{Markus Hennrich~\orcidlink{0000-0003-2955-7980}}
\affiliation{Department of Physics, Stockholm University, SE-106 91 Stockholm, Sweden}

\author{Nikolay V. Vitanov~\orcidlink{0000-0001-6209-547X}}
\affiliation{Center for Quantum Technologies, Department of Physics, St Kliment Ohridski University of Sofia, 5 James Bourchier blvd, 1164 Sofia, Bulgaria}

\title{Motional state analysis of a trapped ion by ultra-narrowband composite pulses}

\date{\today}

\begin{abstract}
In this work, we present a method for measuring the motional state of a two-level system coupled to a harmonic oscillator. Our technique uses ultra-narrowband composite pulses on the blue sideband transition to scan through the populations of the different motional states. Our approach does not assume any previous knowledge of the motional state distribution and is easily implemented. It is applicable both inside and outside of the Lamb-Dicke regime. For higher phonon numbers especially, the composite pulse sequence can be used as a filter for measuring phonon number ranges. We demonstrate this measurement technique using a single trapped ion and show good detection results with the numerically evaluated pulse sequence.
\end{abstract}

\maketitle

The motional state of a trapped ion plays an essential role in most experiments: Not only is it used for quantum information processing \cite{Leibfried2003}, it also has to be very accurately determined for quantum metrology measurements, as it can cause frequency shifts \cite{Rosenband2010, Bergquist2008, Higgins2019}. Due to the importance of motion in trapped ion experiments, various schemes have been developed for motional state detection of the mean phonon number or even the phonon number distribution~\cite{Wineland1987,Meekhof1996,Shen2014, An2014, Um2016,  Ding2017, Ohira2019, Meir2017,Shiqian2017, mallweger2023}. However, most detection methods assume prior knowledge about the initial phonon distribution. Alternatively, one can utilize electron shelving and quantum state engineering techniques to fully determine the quantum state of the ion \cite{Leibfried1997}. For the majority of detection schemes, which rely on coupling individual phonon number changing transitions, extending the technique to higher phonon numbers can be difficult as changes in the phonon number dependent coupling strength become harder to distinguish. 

Composite pulses allow for the creation of a narrow excitation profile where the coupling to phonon-number changing transitions only occurs in a narrow band of motional energies \cite{Torosov2015}. This method was originally developed for the study of nuclear magnetic resonances \cite{LEVITT198661}. It can be implemented to measure the motional state of the ion and can be easily extended to higher phonon numbers since it facilitates the filtering of any detection event outside of a chosen fixed band of motional energies. Composite pulses have been successfully implemented in quantum optics \cite{Schraft2013, Genov2014} and quantum information \cite{piltz2013,schmidt-kaler2003,poschinger2012}.

In this work, we present how the motional state of a laser-cooled ion can be measured independently of the distribution of the motional quanta, using a composite pulse sequence. We not only test the composite pulse method for low phonon numbers, but also investigate a regime with hundreds of phonons.  Our method can in principle be applied both in and outside the Lamb-Dicke regime, assuming the blue-sideband (BSB) Rabi frequencies are well separated. To get a full characterization of the motional state of the ion we propose a sequence of single-shot measurements where different phonon numbers are probed. This technique offers the advantage to provide a characterization of the motional state population within one experimental run, without the need to re-initialize the state. 

The measurements were performed on two different experimental setups both using a single ion trapped in a linear Paul trap, one using a $\mathrm{^{88}Sr^+}$ ion and the other a $\mathrm{^{40}Ca^+}$ ion. The method is not restricted to trapped ion experiments, but works analogously in any other two-level system coupled to a harmonic oscillator where the phonon number solely affects the coupling strength, but not the coupling frequency of the transition.

\begin{figure}[tb]
\includegraphics[width=\columnwidth]{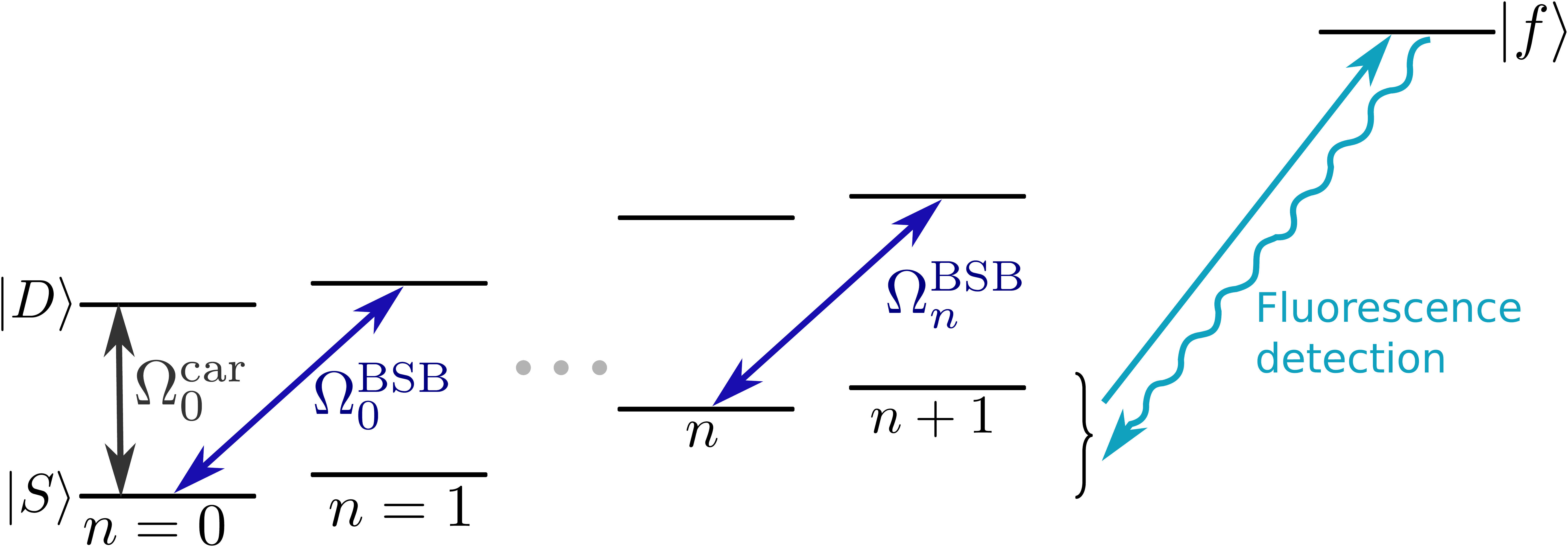}
\caption{The level scheme used for our measurements consists of two electronic states, with the ground state $\ket{S}$ and the metastable excited state $\ket{D}$, coupled to motional states with phonon number $n$. The coupling strength $\Omega_0^{\mathrm{car}}$ of the carrier transition does not depend strongly on the phonon number, whereas the coupling strength of the blue sideband transition (BSB), $\Omega_n^{\mathrm{BSB}} $, scales dependent on $n$. For measurements of the electronic state, fluorescence detection via a shortlived third level is being used.
}
\label{fig:BSB_scheme}
\end{figure}
The combination of a two-level system, composed of a ground state $\ket*{S}$ and an excited state $\ket*{D}$, with a quantum harmonic oscillator generates a ladder-like structure, as illustrated in Fig.~\ref{fig:BSB_scheme}. Blue- and red-sideband transitions (BSB and RSB), detuned from the two level system by the harmonic oscillator frequency can be used to add or remove a phonon between different levels making it possible to move up or down the ladder structure in a controlled way. For the case of trapped ions, the two level system is realized by two (meta-)stable states and the harmonic oscillator is given by one of the motional modes. The BSB coupling strength is given by
\begin{align}\label{eq:coupling}
\begin{split}
       \Omega_n^{\mathrm{BSB}} & =\Omega_0^{\mathrm{car}}e^{-\frac{\eta^2}{2}}\eta\sqrt{\frac{1}{n+1}}L^1_n\left(\eta^2\right)\\ & \approx\eta\Omega_0^{\mathrm{car}}\sqrt{n+1},
\end{split}
\end{align}
where $\Omega_0^{\mathrm{car}}$ denotes the coupling strength of the ground state transition $\ket*{S,0}\leftrightarrow\ket*{D,0}$, $\eta$ is the Lamb-Dicke parameter and $L^1_n$ labels the Laguerre polynomial. Within the Lamb-Dicke regime, meaning for lower phonon numbers, the approximation becomes valid.

\begin{figure}
\centering
    \includegraphics[width=\columnwidth]{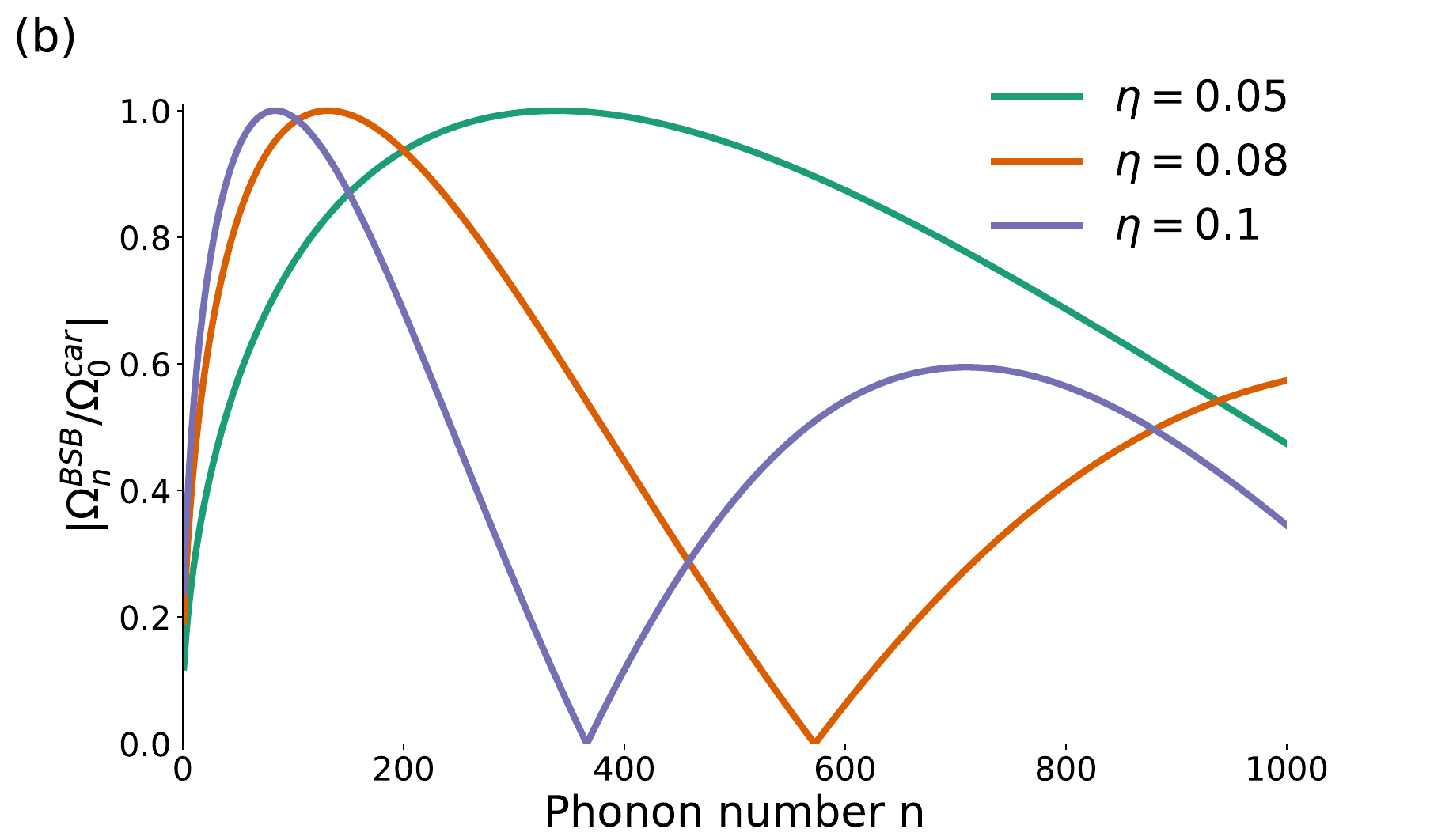}\label{fig:coupling}
    \includegraphics[width=\columnwidth]{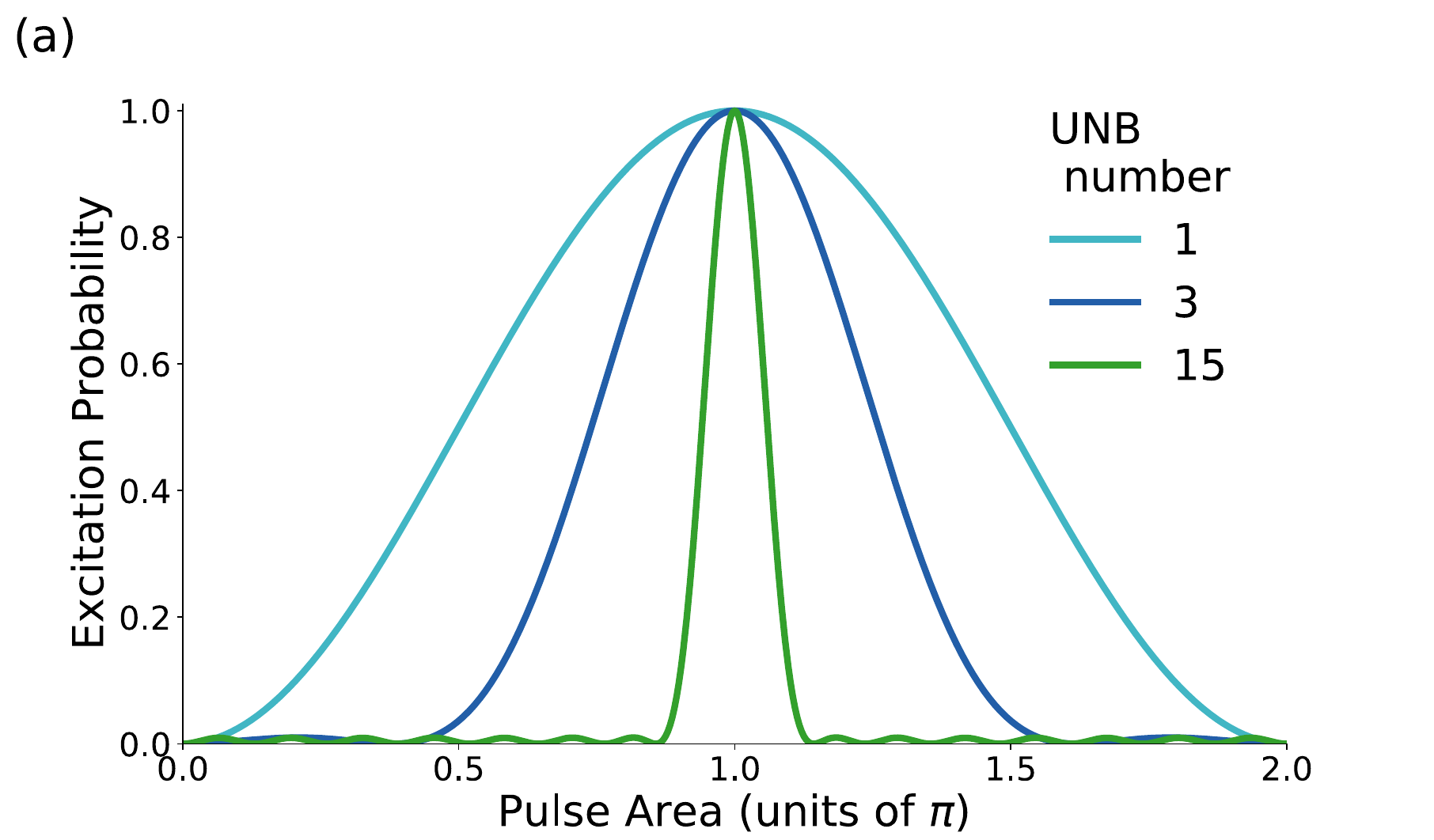}\label{fig:pulse_area}
    \includegraphics[width=\columnwidth]{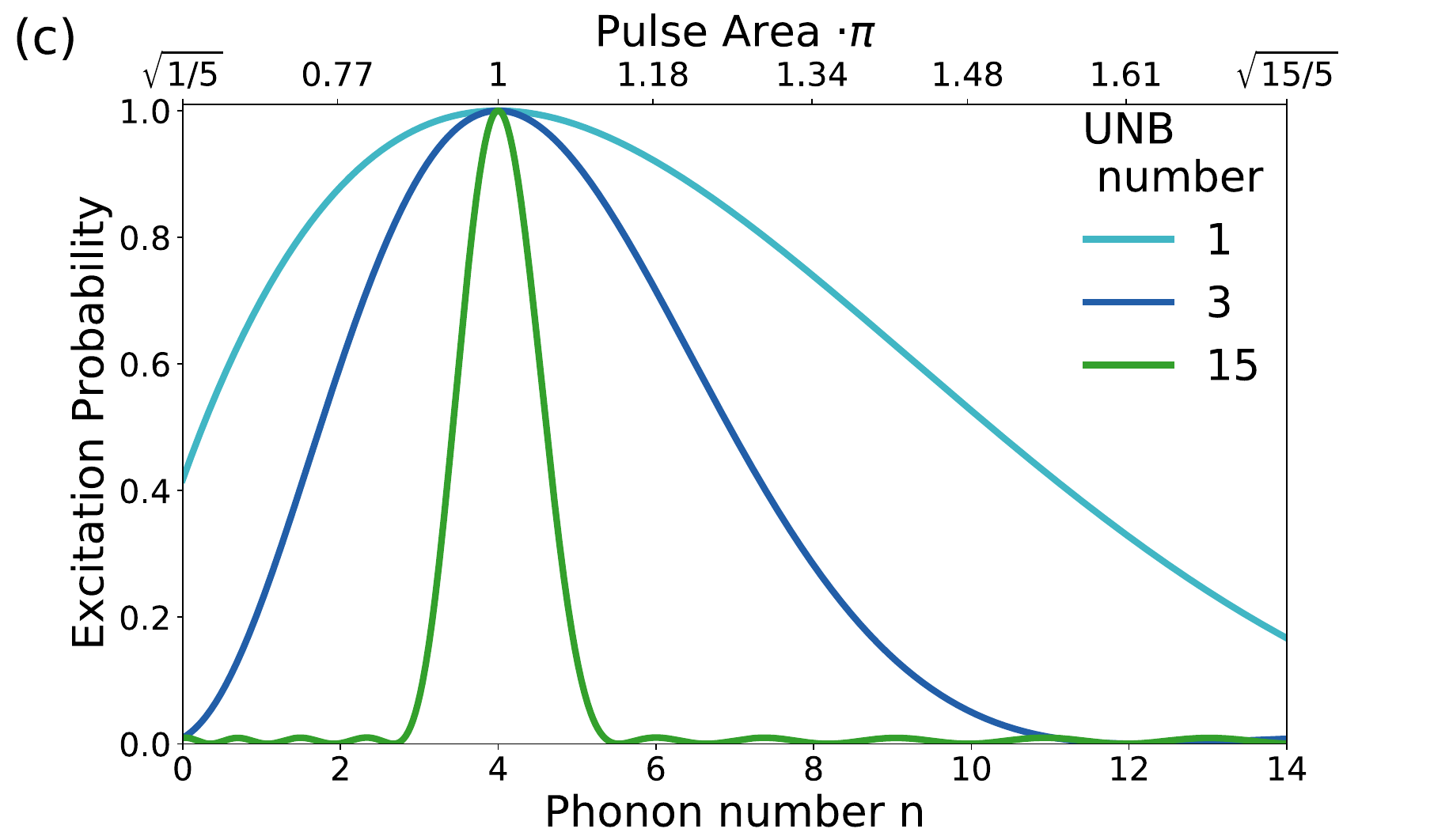}\label{fig:phonons_su}
    \includegraphics[width=\columnwidth]{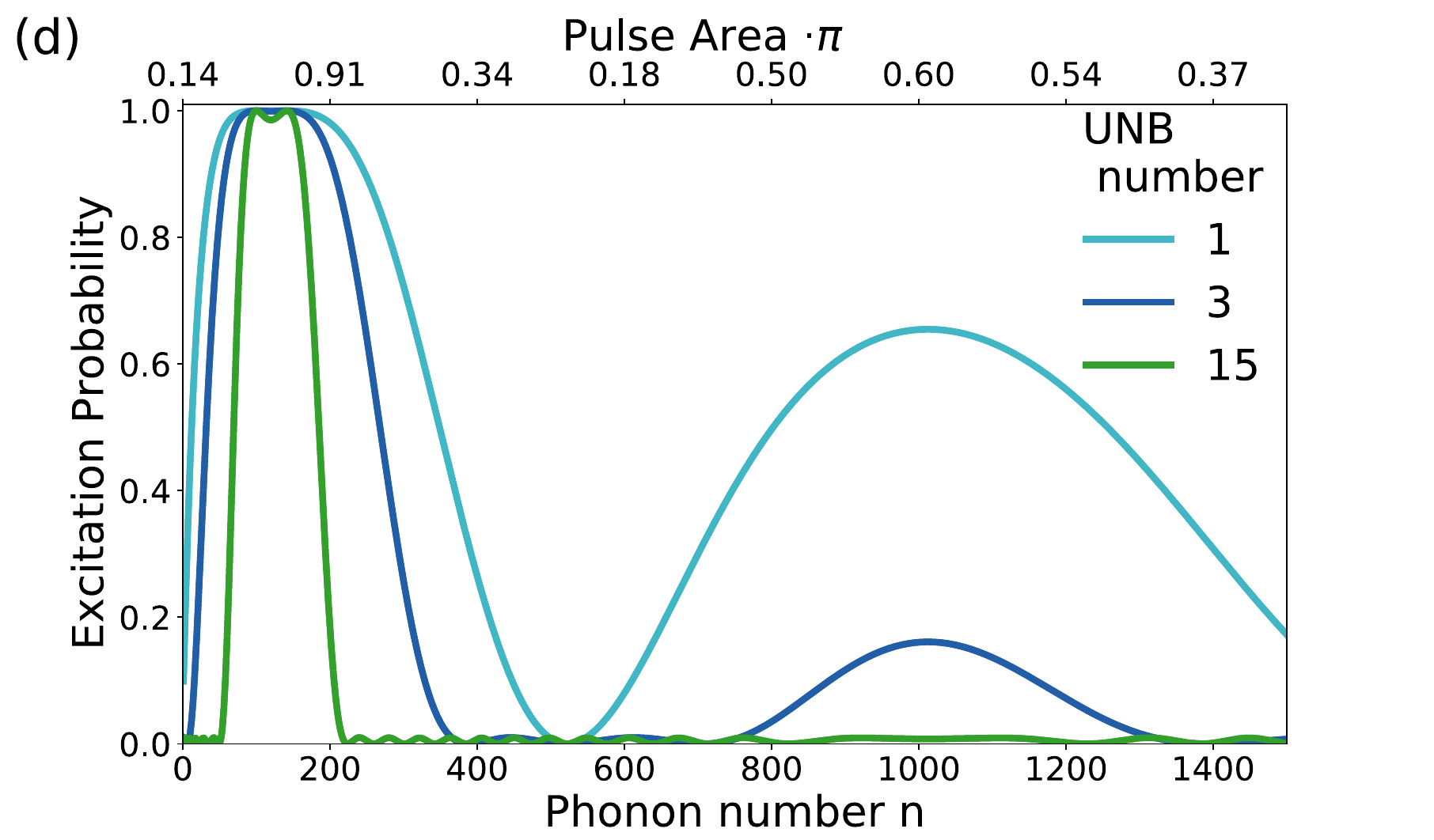} \label{fig:phonons_ibk}
\caption{Composite pulses allow for thermometry on trapped ions. (a) The coupling strength of the BSB transition is dependent on the phonon number and the Lamb-Dicke parameter $\eta$. (b) The pulse area gets narrower with an increasing number of composite pulses with optimized phases. (c) Composite pulse sequence applied to detection of a 4-phonon Fock state. (d) An increasing number of composite pulses also shows an selective excitation profile for $n=100$ phonons.
}
\label{fig:phonons}
\end{figure}

Depending on the Lamb-Dicke parameter, the phonon number and the initial coupling strength, $\Omega_n^{\mathrm{BSB}}$ can vary significantly between adjacent BSB transitions. For lower $n$, the increase of the coupling strength is usually more significant than for high phonon numbers. This can be seen at the slope of the coupling strength of the BSB in Fig.~\ref{fig:phonons}(a). The steeper the slope, the easier it is to resolve a specific BSB transition. However, the opposite is also true; for cases where the slope is less pronounced, a single $\pi$-pulse on the BSB transition also excites neighbouring BSB transitions. The implementation of composites pulses offers a solution to this problem. 

The main idea of the ultra-narrowband (UN) composite pulse method is to use a sequence of pulses to perform an effective $\pi$-pulse on a single BSB transition \cite{Torosov2015}. By increasing the number of UN pulses, the excitation profile can be narrowed. This, however, comes at the cost of a longer pulse sequence which may suffer a reduced effectiveness depending on the coherence time of the experimental system. The UN excitation profile for different numbers of pulses is shown in Fig.~\ref{fig:phonons}, and the individual phase values can be found in the appendix.
Figures~\ref{fig:phonons}(b) and (c) show an explicit simulated example of the excitation profile for different UN sets: For Fig.~\ref{fig:phonons}(c), the ion is assumed to be initialized in state $\ket{S,4}$ before applying the composite pulse sequence for detection of $n=4$. The lower horizontal axis shows the phonon number $n$, the horizontal axis at the top gives the values of the pulse area in units of $\pi$. A pulse area of $\pi$ hence represents a total flip from $\ket{S,n}$ to $\ket{D,n+1}$. In Fig.~\ref{fig:phonons}(d) for higher phonon numbers ($n$=100) the excitation probability as a function of the phonon number is presented for different probing schemes. The figure shows how the excitation profile is narrowed with the implementation of more composite pulses in the sequence, thus producing a more precise filter for motional states outside of the specified range.  

We demonstrate the effectiveness of the technique for lower phonon number measurements, in the range of $0\leq n<10$ using UN composite pulse sequences with a single $\mathrm{^{88}Sr^+}$ ion in a linear Paul trap \cite{Higgins2019}. In this experimental system, we use the $\ket{S}\equiv\ket{5{}^2S_{1/2}\textbf{ }, m_J=-\frac{1}{2}}\leftrightarrow\ket{D} \equiv \ket{4{}^2D_{5/2}\textbf{ }, m_J=-\frac{5}{2}}$ transition as electronic two-level system. The state readout of the two electronic levels is done via state-dependent fluorescence detection using the $5{}^2S_{1/2}\leftrightarrow5{}^2P_{1/2}$ transition. The motional degrees of freedom of the ion in the trap behave like a linear quantum harmonic oscillator. We couple the internal electronic states of the ion with the motional modes by using external laser fields. This results in BSB with the same resonance frequency but different Rabi frequencies, as needed for the UN pulses. The phonon state preparation and UN composite pulse detection sequence is performed on one of the radial modes.\\

Before detection of the phonon number $n$, we first cool the trapped $\mathrm{^{88}Sr^+}$ ion close to its motional ground state using Doppler and sideband cooling, initializing the ion in $\ket{S,0}$. To add a motional quantum, we drive a $\pi$-pulse on the blue sideband (BSB) transition from $\ket{S,0}$ to the excited state $\ket{D,1}$, followed by a  $\pi$-pulse on the carrier transition, as described in \cite{Meekhof1996}. This brings the ion to the state $\ket{S,1}$. Repeating the sequences $m$ times will result in a phonon number increase of $m$. For the detection of $n$, BSB pulses given by the UN pulse sequence are applied. \\
During the fluorescence detection there are then two possible outcomes: Photon scattering during fluorescence detection changes the motional state, however, given that the motional state is unaltered by a positive result, the sequence also allows for non-demolition measurements of Fock states. Note that after this step, the phonon number will be increased by one motional quanta.\\

\begin{figure}[t]
\centering
\includegraphics[width=\columnwidth]{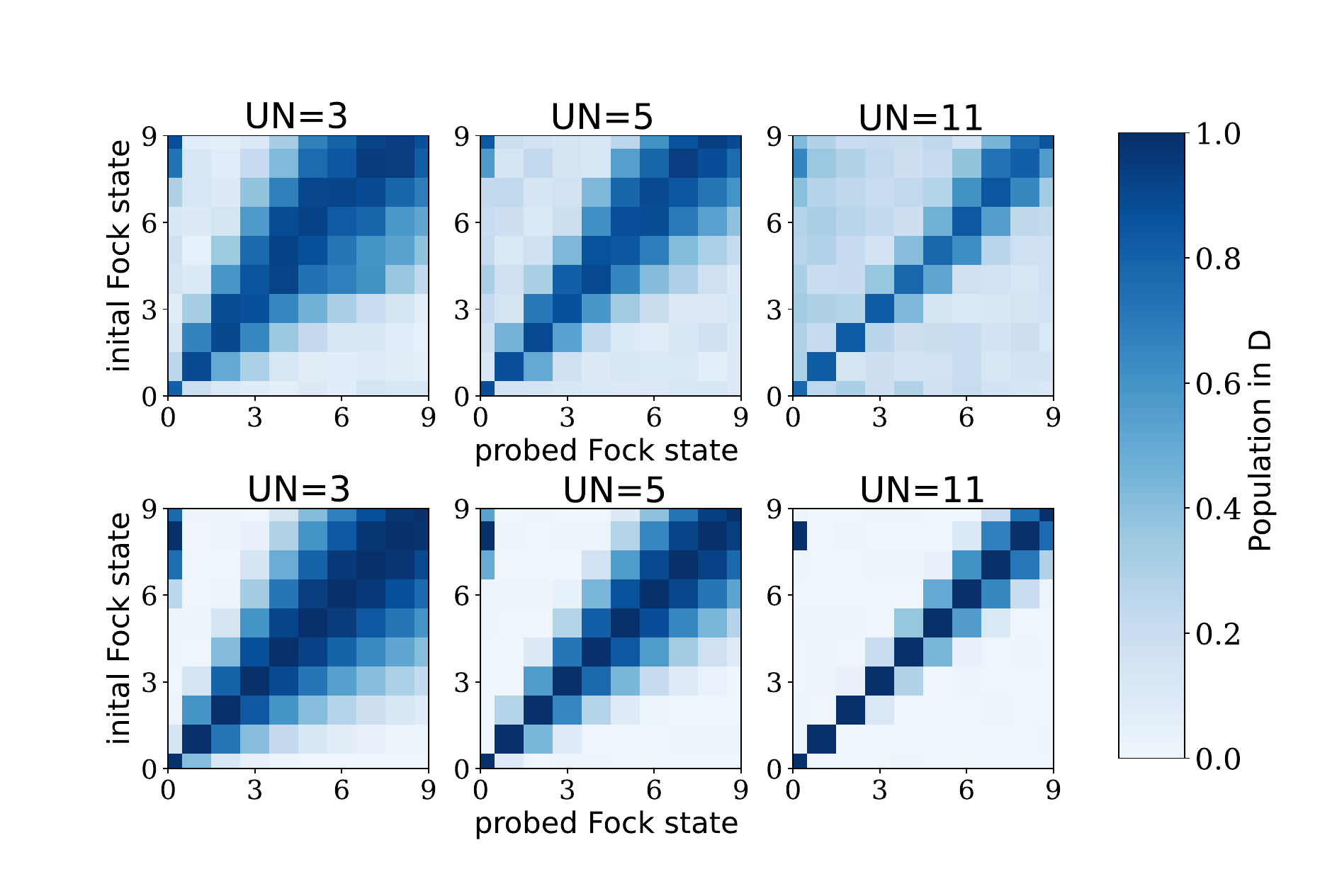}
\caption{Motional state detection using composite pulses.The top row shows our experimental results, which are in good agreement with the theoretical simulation in the bottom row. Using BSB pulses the ion was initialized in different Fock states (vertical axis) and afterwards the different motional states were probed using UN pulses (horizontal axis). The off diagonal excitation decreases with a higher number of UN pulses, leading to more precise results of motional state detection. }
\label{fig:results_su}
\end{figure}

\begin{figure}[t]
\centering
\includegraphics[width=\columnwidth]{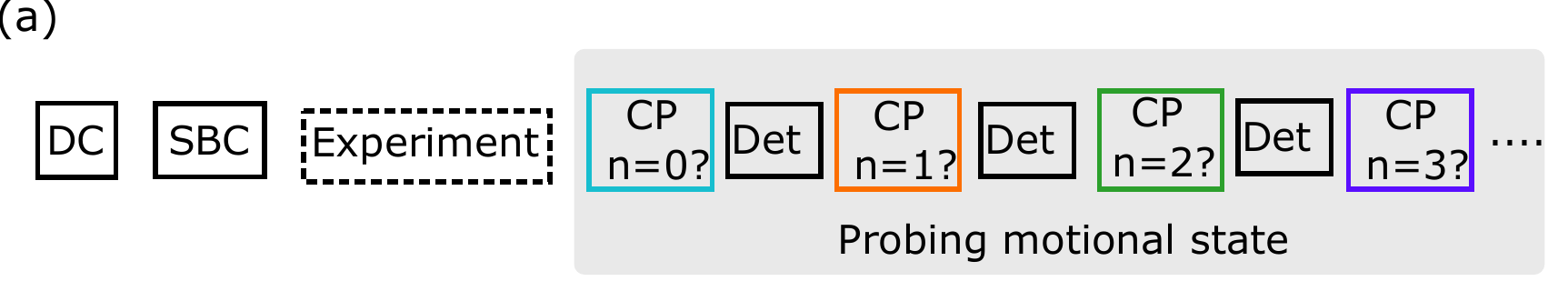}\label{fig:probingscheme}
\includegraphics[width=\columnwidth]{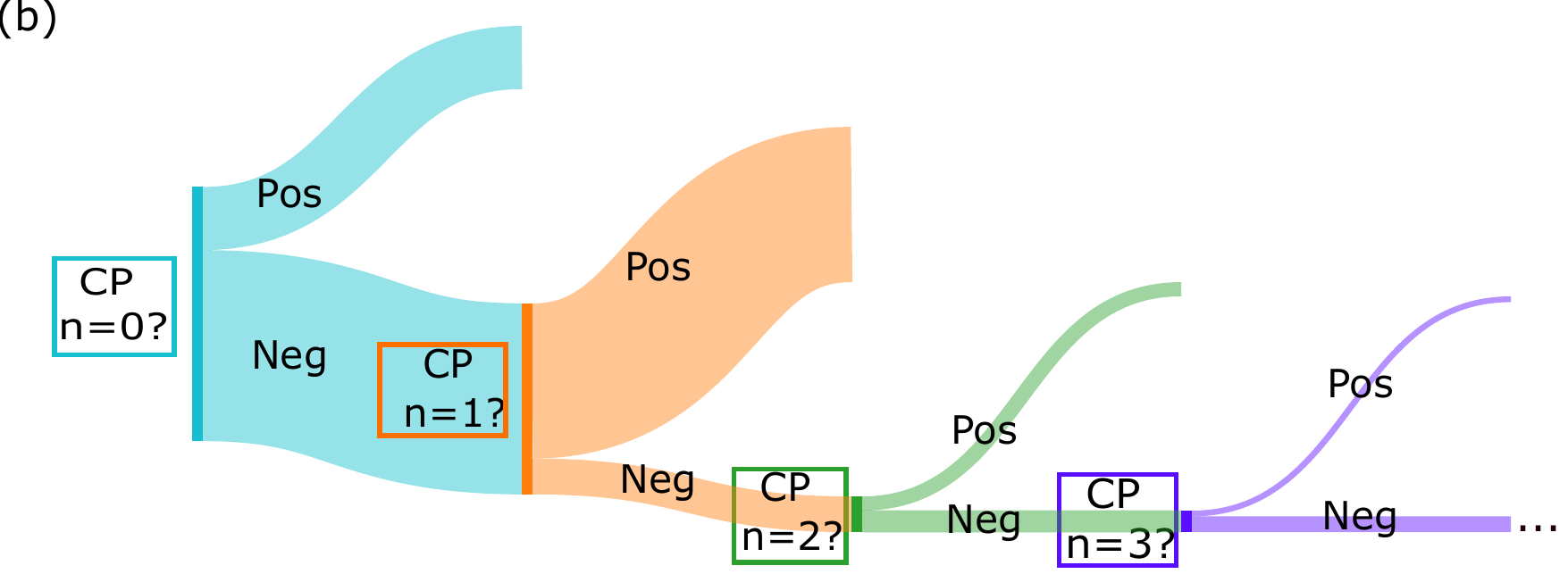}\label{fig:sankey}
\includegraphics[width=\columnwidth]{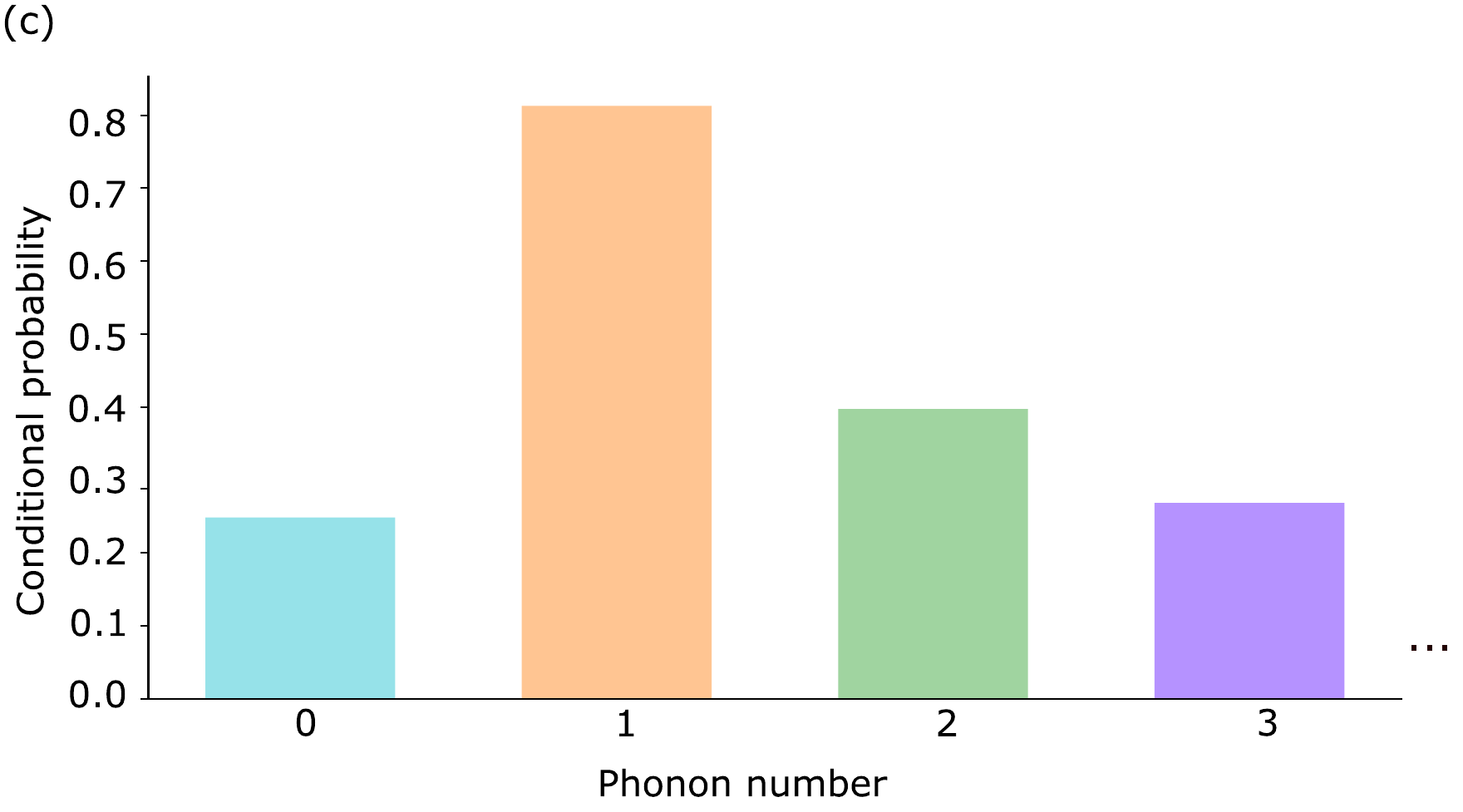}\label{fig:barplot}
\caption{Single-shot measurement using composite pulses. (a) The probing scheme starts by Doppler cooling and sideband cooling (DC and SBC), followed by a chosen experiment. To probe the motional state of the ion a series of composite pulses (CP) and detection (Det) pulses are performed probing different phonon numbers ($n$). If the result of a probing pair for a given phonon number is positive (Pos) the measurement stops, if it is negative (Neg) the next phonon number is probed. (b) A Sankey diagram shows an example of the outcome of a measurement set where the ion was prepared with a phonon number equal to 1. As expected, the majority of positive events come after the probing pair for $n$=1. Positive events from the other probing pairs can be explained by noise and excitation due to other adjacent transitions. (c) The bar diagram gives a conditional probability for the test of this phonon number gives that all previous phonon numbers were tested negative.
}
\label{fig:singleshot}
\end{figure}

For the low phonon number regime we prepared Fock states $\ket*{m}$ from $m=0$ to $m=9$ and probed Fock states $\ket*{n}$ from $n=0$ to $n=9$ using the UN pulse sequence.

The absence of fluorescence indicates $n=m$ leaving the ion in the state $\ket{D, n+1}$. After the state detection was performed the ion was again cooled and initialized in the desired state.\\
The measurements were done for multiple UN pulse sequences consisting of increasing pulse numbers. The results are shown in Fig.~\ref{fig:results_su}. Increasing the number of pulses increases the precision of the detection. The experimental data is in good agreement with the theoretical prediction, described in the appendix.
The increased background of the experimental data can be explained by phase and intensity fluctuations of the laser and excitation due to other adjacent transitions. The composite pulse scheme shows a high detection efficiency for $n=m$. Using the non-demolition technique, it is possible to efficiently verify and continue using the desired motional state in subsequent measurements. For the case $n\neq m$ a higher number of pulses leads to a narrower excitation profile and therefore less excitation of neighbouring BSB transitions, which can be clearly seen. Especially for UN=11 excitations for $n\neq m$ remain negligible for low $n$.

These methods can be used to measure the motional state in the Fock basis in a single experiment run. This involves conducting a series of yes/no tests, successively testing whether or not $n = 0$, $n = 1$,  $n = 2$, etc.
These tests should be set up such that the negative outcomes are non-disruptive (absence of fluorecence light), to allow further tests to be conducted.
A sequence involving successive tests from $\langle n \rangle = 0$ to $\langle n \rangle = 3$ is presented in Fig.~\ref{fig:singleshot}(a).
This was experimentally implemented, and used to probe a trapped ion which was (nominally) prepared in the $n=1$ state, the results are shown in Figs.~\ref{fig:singleshot}(b) and (c).
Experimental imperfections cause the false positives (positive outcomes to the $\langle n \rangle = 0$, $\langle n \rangle = 2$, $\langle n \rangle = 3$ tests) and the false negatives (negative outcomes to the $\langle n \rangle = 1$ test).
This method increases the amount of information about the motional state that can be gained from a single experiment run, which may be particularly useful in experiments with relatively low repetition rates, such as atom-ion collision experiments \cite{trimby2022buffer, feldker2020buffer, mohammadi2021life}. The method can be also implemented as an alternative way to characterize non-thermal motional state distribution after laser cooling necessary for the estimation of secular motion time-dilation shifts in optical clocks \cite{chen2017sympathetic,chou2010frequency,huntemann2016single}.

To demonstrate the application of the UN composite pulses technique in the regime of hundreds of phonons, a single $\mathrm{^{40}Ca^+}$ ion in a linear Paul trap \cite{Guggemos_2015} was used. In this experimental system  we use the $\ket{S} \equiv \ket*{4{}^2S_{1/2}\textbf{ } m_J=\frac{1}{2}}\leftrightarrow\ket{D} \equiv \ket*{3{}^2D_{5/2}\textbf{ } m_J=\frac{3}{2}}$ transition as electronic two-level system. The detection of the ion is done via state-dependent fluorescence detection using the $4{}^2S_{1/2}\textbf{}\leftrightarrow4{}^2P_{1/2}\textbf{ }$ transition. As for the case of the strontium ion, the motional degrees of freedom of the calcium ion in the trap behave like a quantum harmonic oscillator, allowing the coupling of the internal electronic states of the ion with the motional modes using external laser fields. For this measurement set the phonon adding and UN composite pulse sequences are performed on the axial mode. 

We start by cooling the trapped $\mathrm{^{40}Ca^+}$ ion close to its motional ground state using Doppler and sideband cooling, initializing the ion in the $\ket{S,0}$ ground-state. To add the desired motional quanta, a weak electric field with a tunable frequency is applied by means of an electrode situated at the bottom of the trap to excite the axial motional mode of the ion. By controlling the pulse duration and intensity of the ``tickle-voltage'' and using a frequency slightly detuned ($\approx \SI{600}{\hertz}$) from the resonance with the motional mode, the ion can be coherently excited. The pulse length of the tickle voltage is scanned and thus the phonon number is varied. To access different motional energy ranges the axial confinement of the trap is modified by changing the voltage of the end-cap electrodes. 

For the detection of the phonon number in the desired range we implemented a triple detection scheme: Three sets of UN composite pulses followed by a detection pulse are applied. The first set is composed by 5 composite pulses on the BSB, if the motional energy of the ion is in the desired phonon range, this pulse sequence transfers the population from $\ket{S}$ to $\ket{D}$ and a ``dark'' event is detected during the fluorescence excitation. The second set is composed by 3 composite carrier pulses, used to depump any false positive event due to heating rate, in this case a positive event would be detected as ``dark'' and a false one as ``bright''. Finally a third set of composite pulses on the BSB is applied to transfer the population back to the S state. When the energy of the ion is in the selected range the result of the measurement is: ``dark'', ``dark'', and ``bright''. The implementation of this triple detection scheme ensures a suppression of any excitation probability outside of the desired range to about $10^{-4}$. For reference, if only one set of UN composite pulses is implemented on the BSB the excitation probability outside of the desired range is suppressed to about $10^{-2}$.

\begin{figure}[t]
\centering
\includegraphics[width=\columnwidth]{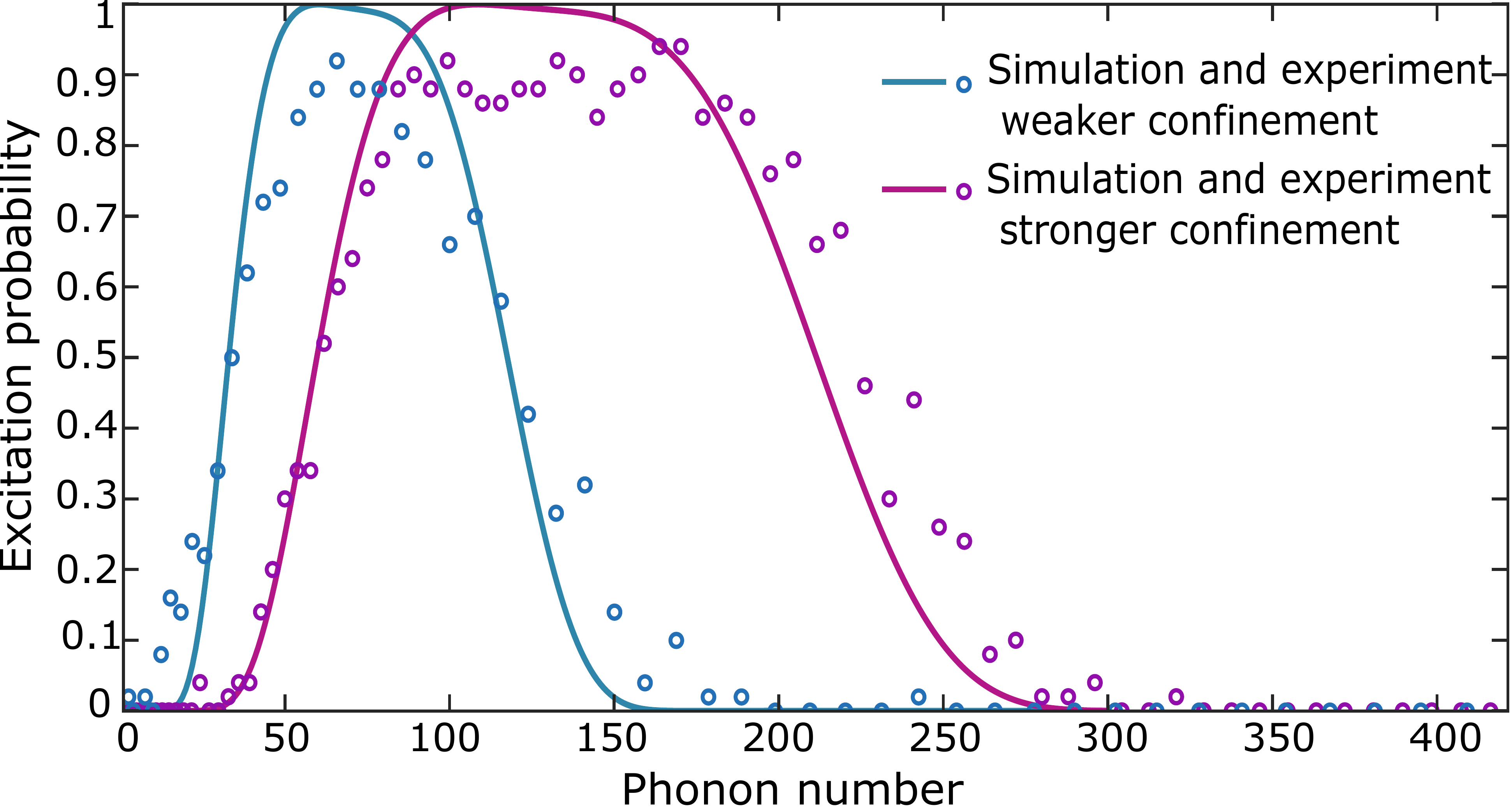}
\caption{By scanning the pulse length of the "tickle voltage" using a slightly detuned frequency ($\approx \SI{600}{\hertz}$) the mean phonon number is "scanned". The excitation probability using the triple detection scheme is measured as a function of the mean phonon number. The measured data for two
different axial confinements (blue circles for $\SI{742}{\kilo\hertz}$ and purple for $\SI{1.329}{\mega\hertz}$) matches closely the predicted theoretical calculations (blue
and pink solid lines).}
\label{fig:results_ibk}
\end{figure}

As illustrated in Fig.~\ref{fig:results_ibk} the technique is successfully implemented to detect two distinct energy ranges, using an axial frequency of $\SI{742}{\kilo\hertz}$ and $\SI{1.329}{\mega\hertz}$. The measured excitation probability as a function of the mean phonon number for both energy ranges is in close agreement with the theoretical estimations. As can be seen in the figure, only energies
in the defined energy ranges are detected as ``positive'' events by the triple detection method: between 35 and 119 phonons for the weak trap setting with $\SI{742}{\kilo\hertz}$ axial confinement and between 63 and 213 phonons for the strong trap at $\SI{1.329}{\mega\hertz}$. As can be seen in Fig.~\ref{fig:results_ibk}, the excitation of the measured data points does not reach the maximum value, this effect can be explained on one side by decoherence processes (electric or magnetic field noise) during the 13 pulses, or it can also be explained by imperfections on the coherent excitation of the ion by means of the ``tickle-voltage''.

In this work we have presented a method for measuring the motional state of a trapped ion using suitable, ultra-narrowband, composite pulses. Our approach is based on a systematic scan of the populations of the phonon state or range. Our method can be seen as an alternative to other traditional methods. It works for $\eta \lesssim 0.3$ and hundreds of phonons. The technique has the advantage that it can be implemented in a non-demolition manner facilitating the design of single-shot measurement schemes to fully probe the motional state of the ion after any given operation. We have presented the implementation of a scheme that probes in a single run the occupation of different phonon numbers.  The measurements performed with single trapped ions of different species show good agreement with the numerically prediction using UN pulses, in the regime of low as well as high phonon numbers. An increase in pulse number also results in a higher accuracy of determining the motional state but comes at a cost of a longer pulse sequences which might exceed the coherence time.

While other techniques will give a more accurate result of the phonon state, especially in the lower phonon number regime, the composite pulse method can be easily implemented in any two-level system coupled to the harmonic oscillator of interest. This might be particularly useful if one needs a lightweight and easy to implement method to test if a single Fock state is populated. For higher phonon numbers it can be used as a filter to determine if the phonon number is in the desired range or not.

The experimental work was also supported by the Knut \& Alice Wallenberg Foundation through the Wallenberg Centre for Quantum Technology [WACQT], by the Swedish Research Council (Grant No. 2017-04638, 2020-00381 and 2021-05811), by the Carl Trygger Foundation, by the Olle Engkvist Foundation, and by the Bulgarian national plan for recovery and resilience, contract BG-RRP-2.004-0008-C01 (SUMMIT), project number 3.1.4 (AQOT). This project has also received funding from the European Union’s Horizon Europe research and innovation program under Grant Agreement No. 101046968 (BRISQ). We also acknowledge funding from the Institut f\"{u}r Quanteninformation GmbH.

\bibliography{main_axiv}

\clearpage
\section{Supplemental Material}
\subsection{Experimental Setup - Stockholm University}

The measurements for detection of lower phonon numbers were performed with a single trapped $\mathrm{^{88}Sr^+}$ in a linear Paul trap. For each individual experimental run, Doppler cooling was performed on the $5{}^2S_{1/2} \leftrightarrow 5{}^2P_{1/2}$ transition, afterwards the motional ground state was reached by sideband cooling on the radial modes. After the cooling steps, the ion was optically pumped into the state $\ket{S} \equiv 5{}^2S_{1/2} \:, m_J=-\frac{1}{2}$.

\begin{figure}[t]
\centering
\includegraphics[width=\columnwidth]{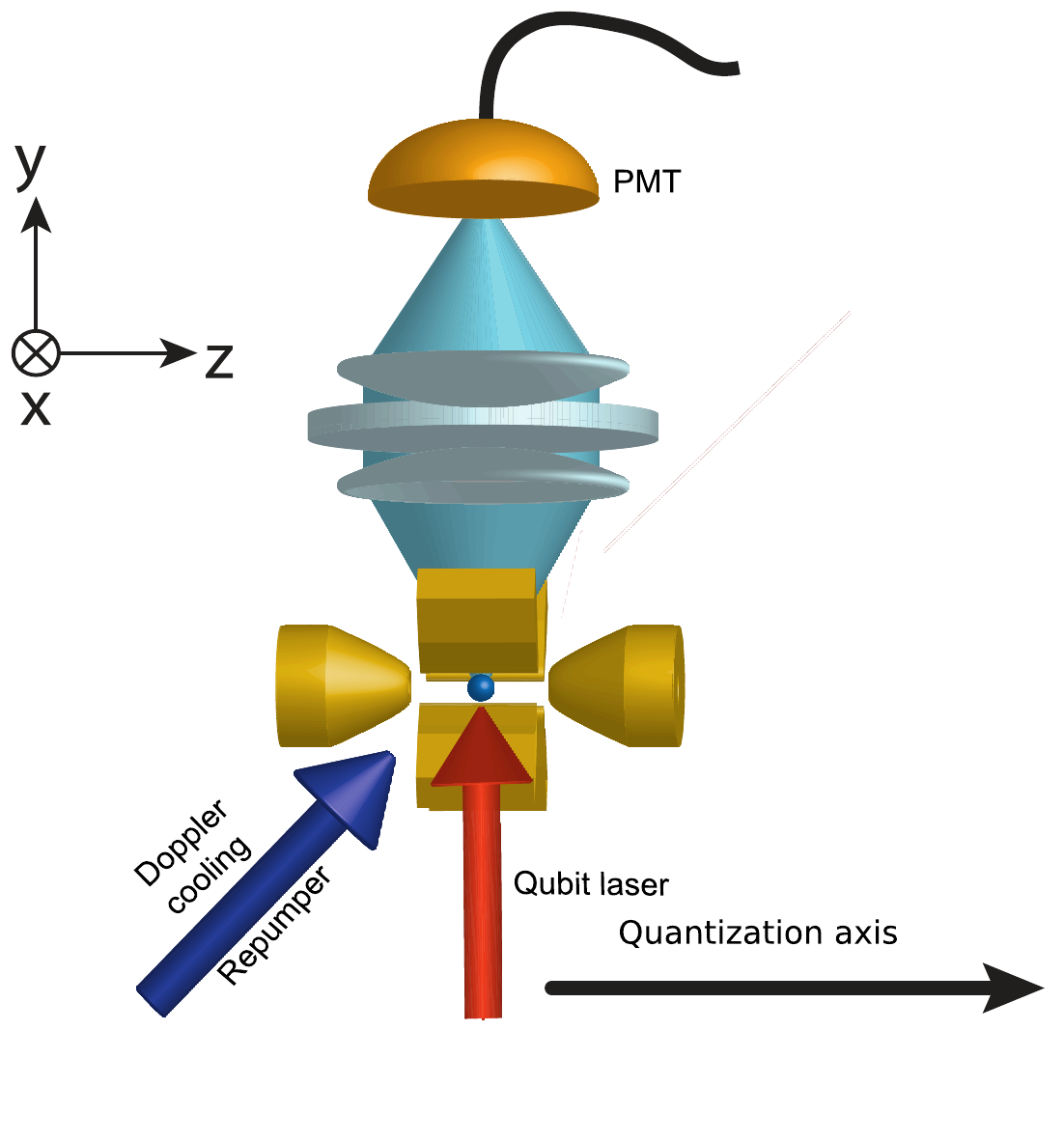}
\caption{The ion is trapped using a linear Paul trap. The cooling and repump lasers are applied $45^{\circ}$ to the trap axis, the qubit laser is applied radially (on the horizontal plane) with $90^{\circ}$ to the trap axis. Fluorescence is detected by collecting light with a large numerical aperture lens and directing it onto a photomultiplier tube (PMT) mounted at the top of the experiment.}
\label{fig:sketch}
\end{figure}

All operations on the electronic two-level transition $\ket{S} \leftrightarrow \ket{D}$, with $\ket{D}\equiv 4{}^2D_{5/2} \:, m_J=-\frac{5}{2}$, were performed with a single beam that is perpendicular to the trap axis and which had a $45^{\circ}$ overlap with each of the two radial modes. Fluorescence detection was performed using a photomultiplier tube (PMT) mounted at the top of the chamber, by excitation of the $5{}^2S_{1/2} \leftrightarrow 5{}^2P_{1/2}$ transition. During Doppler cooling and detection sequences a repumping laser at \SI{1092}{\nano\meter}, driving the transition $4{}^2D_{3/2} \leftrightarrow 5{}^2P_{1/2}$, was also used to remove any population that spontaneously decays into the metastable $4{}^2D_{3/2}$ state. A sketch of the setup is shown in Fig.~\ref{fig:sketch}.

The composite pulse sequences were performed on a blue sideband (BSB) transition on a radial mode of the ion with Lamb-Dicke parameter $\eta=0.036$. The radial sideband was detuned from the carrier transition by $\Delta=2\pi\cdot\SI{1.74}{\mega\hertz}$ and the coupling strength for the transition $\ket{S, n=0} \leftrightarrow \ket{D, n=1}$ was $\Omega_0=2\pi\cdot\SI{5.04(0.03)}{\kilo\hertz}$. The measured heating rate of the radial mode used for the experimental sequence was $13.7(4)\,$quanta/s.
\subsection{Experimental setup - University of Innsbruck}
\begin{figure}[t]
\centering
\includegraphics[width=\columnwidth]{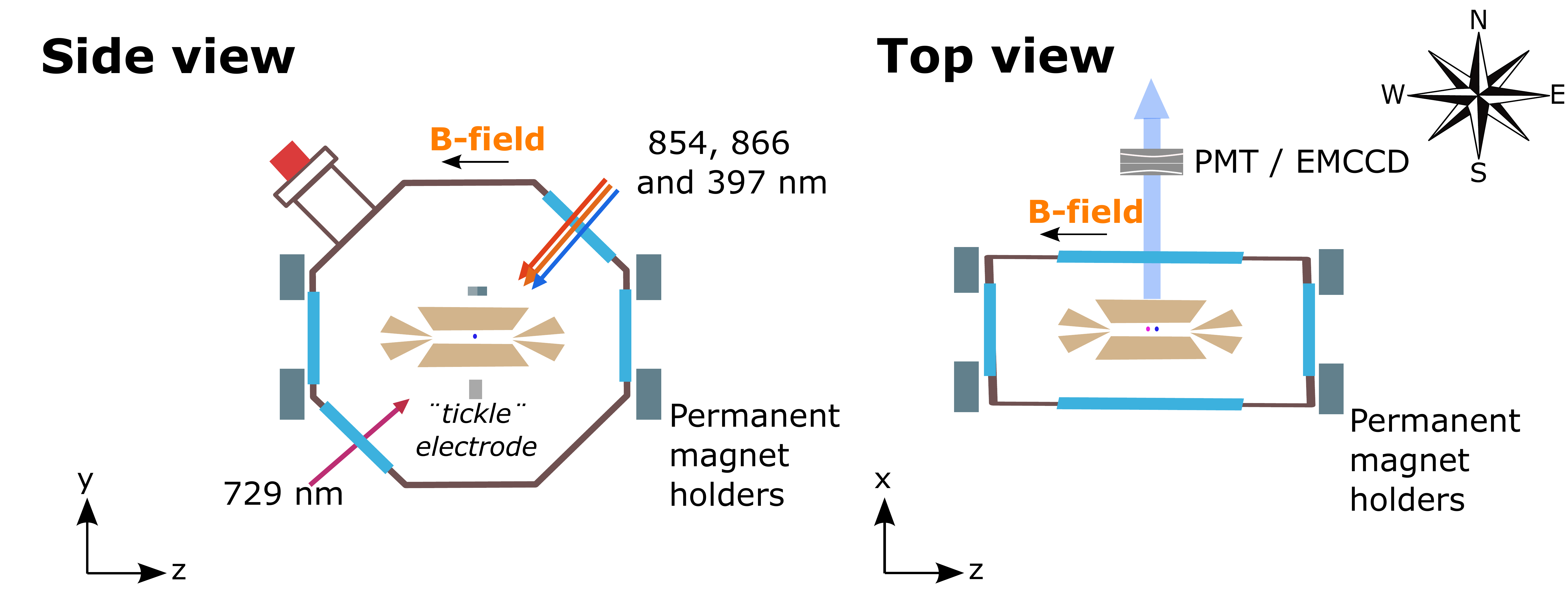}
\caption{A $\mathrm{^{40}Ca^+}$ ion is trapped using a linear Paul trap. The different lasers and imaging directions are illustrated with respect to the cardinal directions (indicated by the compass rose). The magnetic quantization field (B-field) along the z-axis of the trap is provided by a pair of concentric permanent magnet holder situated in the west and east viewports of the chamber. The "tickle" voltage electrode is situated below the trap along the y-axis. The fluorescence emitted by the ion is detected by a photomultiplier tube and an electron multiplying charge-coupled device camera through the northern viewport. }
\label{fig:sketchIBK}
\end{figure}

The measurements for the higher phonon numbers were performed with a single trapped $\mathrm{^{40}Ca^+}$ in a linear Paul trap. The transition $4{}^2S_{1/2} \leftrightarrow 4{}^2P_{1/2}$ at \SI{397}{\nano\meter} is used for Doppler cooling and internal state discrimination in combination with the $3{}^2D_{3/2} \leftrightarrow 4{}^2P_{1/2}$ transition at \SI{866}{\nano\meter}, the $3{}^2D_{5/2} \leftrightarrow 4{}^2P_{3/2}$ at \SI{854}{\nano\meter} is used as a repumper for efficient state preparation. 
The $4{}^2S_{1/2} \leftrightarrow 3{}^2D_{5/2}$ quadrupole transition at 729~nm is used for two-level operations using a single beam orthogonal to the quantization axis. The same transition in combination with the \SI{854}{\nano\meter} is used for sideband cooling. The fluorescence is split and detected using both a photomultiplier tube (PMT) and an electron multiplying charge-couple device (EMCCD) camera through the northern viewport as shown in Fig.~\ref{fig:sketchIBK}.

To add the desired motional quanta to the axial motional mode of the single ion, a weak electric field with a tunable frequency is applied by means of a ``tickle" electrode situated at the bottom of the trap. 

\subsection{Ultra-Narrowband composite pulse sequence}
Table I lists UN composite pulse sequences, which was used for the presented measurements.
A measure of the performance of these sequences is the dimensionless parameter $\alpha$, which measures the width of the excitation profile, outside of which the excitation probability is $\leq 0.01$.
For a single resonant pulse the value of $\alpha$ is $0.936$.

For the first two transitions $\ket{0}\to\ket{1}$ and $\ket{1}\to\ket{2}$ to be separated (with a contribution of less than 1\% from the other transition), we must have $\alpha < \sqrt{2}-1 \approx 0.414$.
As Table I shows, ultra-NB sequences of 5 or more pulses can achieve this separation, meaning that the signal for the first peak $\ket{0}\to\ket{1}$ will be accurate up to 1\% error (the second peak has also a contribution from the third transition $\ket{2}\to\ket{3}$).
If we wish to separate the first three transitions $\ket{0}\to\ket{1}$, $\ket{1}\to\ket{2}$ and  $\ket{2}\to\ket{3}$, we must have $\alpha < \sqrt{3/2}-1 \approx 0.225$.
As Table I shows, ultra-NB sequences of 9 or more pulses can achieve this separation.
In this case the signals from  the first two peaks $\ket{0}\to\ket{1}$ and  $\ket{1}\to\ket{2}$ will be accurate up to 1\% error.
Finally, if we wish to separate the first four transitions, from $\ket{0}\to\ket{1}$ to $\ket{3}\to\ket{4}$, we must have $\alpha < \sqrt{4/3}-1 \approx 0.155$.
As Table I shows, ultra-NB sequences of 13 or more pulses can achieve this separation.\\

\begin{table}[tbph]\label{table:UN}
\begin{tabular}{lll}
\hline
CP & phases $(\phi_1,\phi_2,\ldots,\phi_N)$ & $\alpha$ 
 \\ \hline
$\text{UN}3$ & $(0,0.587,1.174)\pi$ & $0.551$ \\
$\text{UN}5$ & $(0,0.237,1.583,0.929,1.166)\pi$ & $0.361$ \\
$\text{UN}7$ & $(0, 0.299,0.972,0.850,0.727,1.400,1.700)\pi$ & $0.265$ \\
$\text{UN}9$ & $(0,0.392,0.032,1.996,0.597,1.199,1.164,0.805,$ & \\
 & $ 1.198)\pi$ & $0.209$ \\
$\text{UN}11$ & $(0, 0.429, 0.380, 0.931, 1.285, 0.991, 0.695, 1.047,$ & \\
 & $ 1.598, 1.551, 1.982)\pi$ & $0.172$ \\
$\text{UN}13$ & $(0, 0.182, 1.949, 1.435, 1.008, 1.424, 1.256, 1.084,$ & \\
 & $1.491, 1.057, 0.544, 0.321, 0.514)\pi$ & $0.146$ \\
$\text{UN}15$ & $(0, 0.021, 0.512, 0.894, 0.841, 0.475, 0.940, 1.059,$ & \\
 & $1.166, 1.625, 1.264, 1.212, 1.604, 0.095,0.101)\pi$ & $0.127$ \\
\hline\\
\end{tabular}
\caption{UN composite sequences which produce significant excitation inside the range $[\pi(1-\alpha),\pi(1+\alpha)]$, and negligible excitation below 0.01\% outside this range.}
\end{table}

\end{document}